# A Design Tool to Reason about Ambient Assisted Living Systems

Hong Sun, Vincenzo De Florio, Chris Blondia
*University of Antwerp*
*Department of Mathematics and Computer Science*
*Performance Analysis of Telecommunication Systems group*
*Middelheimlaan 1, 2020 Antwerp, Belgium, and*
*Interdisciplinary institute for BroadBand Technology*
*Gaston Crommenlaan 8, 9050 Ghent-Ledeberg, Belgium*
*{hong.sun, vincenzo.deflorio, chris.blondia}@ua.ac.be*

## Abstract

*This paper proposes a design tool to investigate the properties and emergent behaviours of a special class of Ambient Assisted Living systems, namely mutual assistance communities where the dwellers contribute to each other's well being. Purpose of our system is to understand how mutual assistance communities work, what consequences a design decision could ultimately bring about, and how to construct care communities providing timely and cost-effective service for elderly and disabled people. We prove that mutual assistance between dwellers can provide care in time, and decrease the requirement for professional medical service. The simulation results show that with the existing rules most of the requirements for help can be solved or promptly initiated inside the community before their members resort to external professionals.*

## 1. Introduction

As well known, the proportion of elderly people kept increasing since the end of last century. The European overview report of Ambient Assisted Living (AAL) investigated this trend [1]. The studies of EUROSTAT [2] indicated that: "The share of the total European population (EU 15) older than 65 is set to increase from 16.3% in 2000 to 22% by 2025 and 27.5% by 2050. The share of the population aged over 80 years (3.6% in 2000) is expected to reach 6% by 2025 and 10% by 2050."

Studies of Counsel and Care in UK find out that these elderly people would prefer to live in their own house rather than in hospitals, thus they need support to remain independent at their home [3]. Figure 1 shows how people's daily lives are hampered by aging

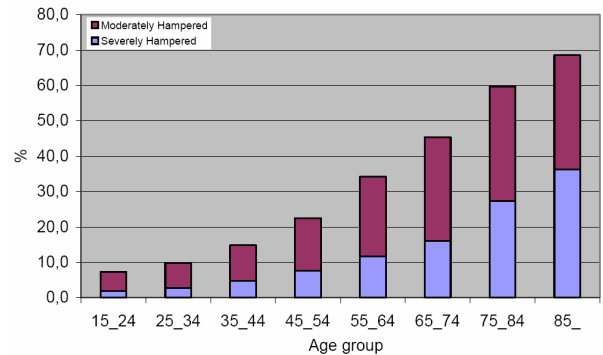

**Fig. 1 Percentage of people hampered in daily activities, by age[2]**

[2]. In order to improve the quality of the life for the elderly and disabled people, it is important to guarantee that assistance to those people should be timely arranged in case of need.

Several researches have been undertaken already to build intelligent care communities based on agents – see for instance [4], [5], and [6]. In particular, in [5] Beer developed his INCA (Integrated Community Care) system with AUML (Agent-oriented Unified Modeling Language), intending to develop care plans for the elderly people. In Beer's system, a coordinator is responsible for preparing the plan, and different care providers supply various services. Non-medical services are carried out in his system by so-called informal carers – e.g. relatives, friends, neighbours etc.

These previous researches are excellent attempts to provide quality of life to the elderly people. In this paper, the system we propose will also contain actors similar to the ones deployed in the previous research. But what is emphasized in this paper is how the system reacts to some abrupt call for help, how to provide these helps timely and cost-effectively and how to increase the degree of satisfaction of the caller. The neighbours will play an important role in this paper. The simulations will show that the system can provide a quick response to the requirement for help from the availability of motivated neighbours.

## 2. The Ambient Assisted Living System

As presented in Sect. I, elderly people need support to maintain their independent life. The call for medical treatment should be provided from the professionals in hospitals, and the call for social activities is normally provided by their relatives, friends, and neighbours. The situation in real-life is that families and friends are not necessarily located nearby, but sometimes far away. In order to provide help timely and cost-effectively, the best solution seems to resort on help from their neighbours. Neighbours are adequate for the task of social activities: they are close to the caller, thus their help could be more timely provided. Furthermore, these interactions could increase the harmony within the community.

Obviously the neighbours may be also not always available, and it is also not an efficient option to call the neighbours one by one to ask for help. An important function of our AAL system is automatically requesting help when the elderly people are in need. Wireless sensors will be installed in the house or even embedded around the body of the elderly people, record their movements and trigger alarm if accidents are detected. Bayesian artificial intelligence tools will be used to capture the behavioural patterns of the user and detect possible anomalies [7]. Besides asking for help, the system will also automatically collect the status of their neighbours, and automatically select one neighbour to provide the service when more than one neighbour is available. The coordinator will record the information of who provided help. The dwellers that provided help could get some reward both morally and financially, at the same time improving their "image" within the community. The willingness of the neighbours to provide help is crucial to the success of this system. The simulation in this paper investigates the satisfaction of the callers with different ratios of callers and help providers in this community.

In this paper, we made the assumption that there will be dwellers in this community who are glad to provide non-medical services to their neighbours. If dwellers ask for medical treatment, the coordinator will forward their requests to hospitals or nursing systems. The investigation in this paper is mainly about non-medical assistance provided by the neighbours.

## 3. Building the AAL System

The system proposed in this paper is built up by an $n \times n$ cell grid; each cell in this grid will represents an individual, and the whole grid will be treated as a community. The individuals in this community will be divided into three types: client, provider, and neutral. Their functions are described in table 1.

| Cell types | Description |
|---|---|
| Client | Client cells represent elderly or disabled people in the community. Their states will turn to neutral after the service completed. |
| Provider | Provider cells represent people who are providing help in the community. After one service completed, their states will either turn to neutral or keep as provider, depending on different help policies. |
| Neutral | Neutral cells represent those people who neither ask for help, nor provider help. They will not take any actions. |

**Table 1. Cell types and their functions**

Fig. 2 shows a 5×5 cell grid. In this grid, cell A has four adjacent neighbours, E, S, W and N, in four directions. Besides these four adjacent neighbours, cell A also has four second-close neighbours, EE, SS, WW and NN. The policy is that if cell A calls for help, it will firstly look for help from its four adjacent neighbours; if none of them could provide help, then A might look for help from its second-close neighbours. For the sake of simplicity we assumed that each individual interacts only with up to 8 neighbouring cells, arranged as in Fig.2

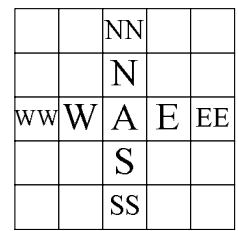

**Fig. 2 Cell A and its neighbours**

Fig. 3 shows the graphic interface of the simulation model. This figure depicts the initialization of a 30×30 cell grid. The cells in blue are neutral cells; the red cells are client cells and the cells in white are the cells providing help. The rates of these three types of cells can be manually set. In Fig. 3, the client cells account for 30.56%; the provider cells account for 34.11% and the neutral cells account for 35.33%. The satisfaction is 0 because there is no service provided in initial.

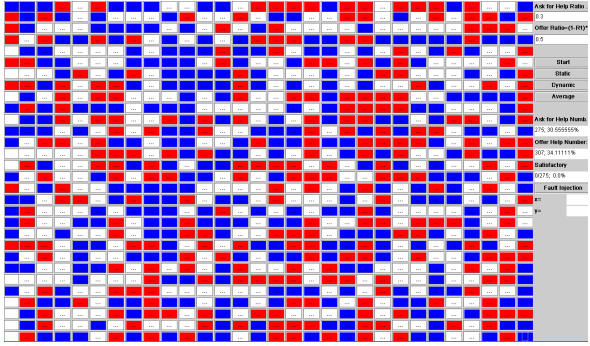
**Fig.3 Initialization of a 30×30 grid cell grid**

Although the cells are divided into different types, the prototype of the cells is homogenous. Every cell in the grid has the same parameters; the different values of these parameters bring the diversity of the cells. There are three parameters that need to be illustrated:

Req_value: This value indicates the amount of work required by a client for one service. For provider or neutral cells, their req_value will be 0. Provider cells will prefer to provide help to client cells with a higher req_value.

Pro_value: This value indicates how fast a provider can perform a service. Client cells will prefer to accept help from provider cells with a higher pro_value.

Work_status: This parameter indicates the status of cells. When a cell is receiving/providing service, its work_status is set to "true"; otherwise, it is set to "false". When the service is finished, work_status is automatically set to false again.

Once a cell grid is initialized, it can start work. The neutral cells keep idle; the client cells call for help and the provider cells provide service to the client cells. The system is run by discrete time steps. During every step, different actions will be taken to send/detect alarm and carry out service. There are various methods to provide service, and one task of this investigation is to find out a good policy to provide help. The policy to provide help that was used in this paper follows the procedures below:

1. Detect: The cells that are able to provide help will look for states of their four adjacent neighbous, detect whether they need help, and provide help to one neighbour if needed;

2. Accept help: The cells that are requiring help will check the states of their neighbours, choose one neighbour to accept the service if available. And the work_status of that cell will change to true.

3. Confirm service: The cells that are offering help will check whether their help offers are accepted and, if that is the case, their work_status will be set to true.

4. Distant help: This is an optional action that could enlarge the contact areas. The cell will iterate the process 2 to 4, but the object is changed from four adjacent negibours to those second close neighbours.

5. Service: The cells that established the client-provider relationship will start the service process.

6. Finish: When the service is finished, the work_status of client and provider will be set to false.

The duration of the service process carried out between a client and provider is regulated by the req_value variable, which gets decreased at each time steps according to the following formula:

$$req\_value = req\_value - \frac{1}{10} pro\_value . \quad (1)$$

During every discrete step in the process of development, the req_value will decrease by ten percent of the pro_value. The steps required to finish one service depend on the amount of the work required and the speed of the service provider. When req_value reaches 0, the process is considered as finished

## 4. Simulation

The simulations of this AAL system are carried out on two directions; these two directions are named as static simulation and dynamic simulation. In the static simulation, the statuses of the cells are set in the initialization. The cells will not change their types except for client cells that received required service and turned to neutral cells. This investigation focuses on how the community reacts to specific situations. In the dynamic simulation, a small portion of the cells will be randomly chosen and change their states randomly during the simulation. This investigation targets on investigating the performance of the system with these dynamic changes. Both of the two simulations are designed to investigate whether the community could meet various rates of service requirement, and aim to find out a policy to deliver the service efficiently.

### 4.1 Static Simulation

In the static simulation, the ratio of the client cells and provider cells are set to various values in initialization. There are two features investigated in this part: The first one is to tell, with different percentages of client cells asking for help and different percentages of provider cells providing help, how many client cells can receive immediate help. The second feature investigated in the static simulation is how many of them could get help at last, i.e. immediately or after some time. Simulations are taken in a 30×30 cell grid.

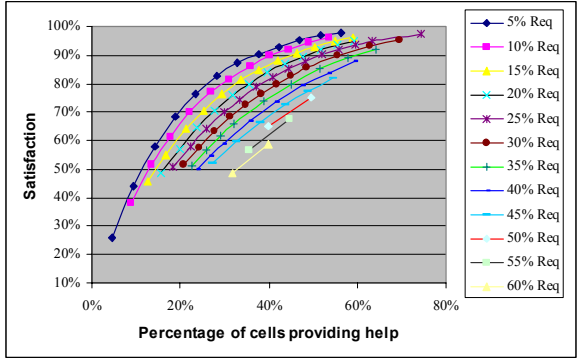

**Fig. 4 Immediate satisfaction**

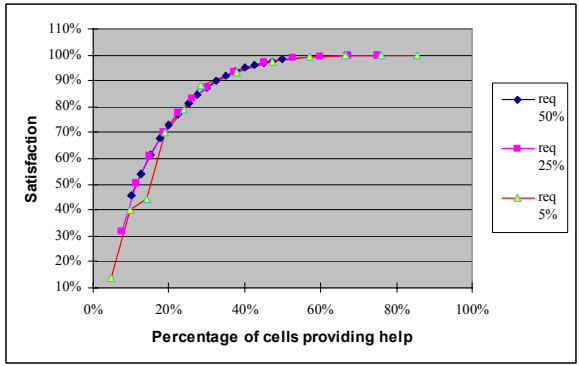

**Fig. 5 Satisfaction after iterations**

Figure 4 shows how many client cells can receive help immediately. The curves with different colours represent the different rates of the client cells in the initialization state, from 5% in the top to 60% in the bottom. The coordinate *x* indicates the percentage of the provider cells in the initial state; the coordinate *y* shows the degree of satisfaction of the client cells, that is, the percentage of the client cells that received help from their neighbours. A provider can serve only one client cell.

Figure 5 shows the simulation results when a provider cell is allowed to provide service again after it finished one service. From Fig. 5, it can be observed that although the ratio of client cells is quite different, their satisfaction is generally the same when given the same percentage of provider cells. In this view, Fig. 5 suggests that if given enough time to provide help, the satisfaction is only decided by the percentage of help providers. The satisfaction of client cells is more than 90% when the help provider rate gets over 40%. When the help provider rate is more than 60%, the satisfaction is almost 100%.

These two figures show that the satisfaction of client cells will increase with a lower percentage of client cells and a higher rate of provider cells. In Fig. 4, when the rate of provider cells is higher than that of the client cells, most of the requests for help will be provided immediately. In Fig. 5, the time limitation is loosed, and almost all the client cells will ultimately receive help.

### 4.2 Dynamic Simulation

The static simulation focused on how the community provides service in some statically defined situation. The result of the static simulation is promising, but does not fully resemble reality. In reality, the community is dynamic and the individuals are changing their states all the time. Thus the dynamic simulation is taken to reflect how the community works dynamically.

Compared with the static simulation, the most significant novelty in the dynamic simulation is that the cells in the grid will randomly change their states. The system will randomly choose some cells in the community and these cells will change their states randomly. The purpose is to investigate whether the community could maintain a balance and provide service continuously and stably.

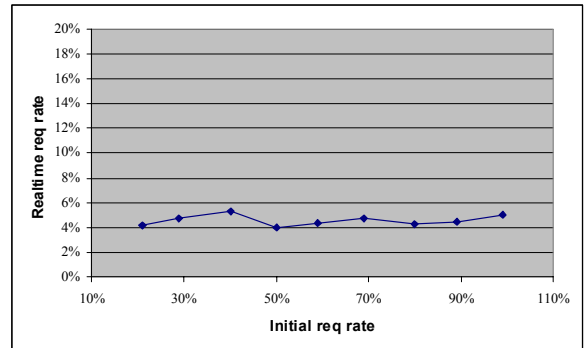

**Fig. 6 Impact of initial state**

The simulation portrayed in Fig. 6 investigates the impact of the initial state. The coordinate *x* shows the rate of the client cells in the initial state, and the coordinate *y* shows how many cells are still requiring help (did not receive service). The community in Fig. 6 is organized as a 30×30 cell grid. After initialization, the community will develop step by step. During every step, there will be one cell randomly chosen to change its state. In this run the chance it changes to a neutral cell is 50%, while the chance it turns to be a client cell or provider cell is both 25%. The system lets the community develop 5000 steps to reach a stable state. After that, the system records periodically the number of cells still asking for help (every 50 steps). This is done for 100 cycles. The average values of the calling for help are shown in the coordinate *y*. Fig 6 shows that the initial state of the community does not take any impact on the community: after the community reaches a stable state, the cells asking for help is around 5%.

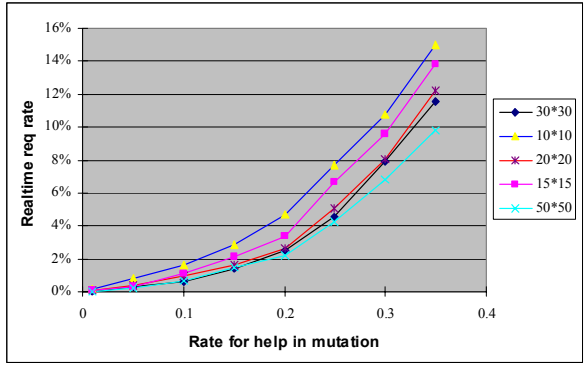

**Fig.7 Real-time satisfaction I**

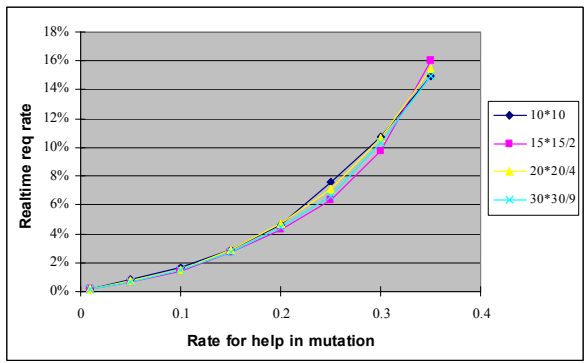

**Fig. 8 Real-time satisfaction II**

Fig. 7 and Fig. 8 show how the dynamic change of roles within the community does influence the overall satisfaction of the community. Clearly this dynamicity reflects the way a real community behaves, with its members changing their willingness to be neutral or provide help, or their need for help.

In both Fig. 7 and Fig. 8, curves with different colours represent different sizes of the cell grid. The coordinate *x* indicates the probability that a changing cell becomes a requestor of help. The probability to become neutral is both 50% in the two figures, hence that of becoming a provider is 50%-*x*. The coordinate *y* shows the rate of the cells that are asking for help. The difference between Fig. 7 and Fig. 8 is that in Fig. 7, only one cell will change state regardless the size of the grid. In Fig. 8, the cells that change their states in every step are 1% of the total cells. That is in a 10×10 cell grid, only one cell change its state every step, and in a 30×30 cell grid, 9 cells change their states.

Figure 7 and Fig. 8 reflect two features of this community: One is that for the dynamic community, the change probabilities of the cells are crucial to the satisfaction of the community: When the probability to become a client cell equals the one to become a provider, the satisfaction is around 94%, i.e. only 6% cells are not served within the community. When the chance to become client cell is lower than 10%, the satisfaction of the community gets over 99%.

The other feature observed is that the size of the cell grid does not bring a noteworthy effect to the satisfaction of the community. Although larger grids have a better performance than smaller grids in Fig. 7, this difference is brought by the different ratio of the cells changing their states in every step. In Fig. 7, the ratio of the cells changing their states in one step is 1/100 and 1/900 for those two cell grids 10×10 and 30×30. When this ratio of the 30×30 cell grid increased to 9/900 in Fig. 8, the satisfaction of the community drops down and gets close to that of the 10×10 grid. This phenomenon also suggests that the lower the ratio of the cells changing their states, the higher the satisfaction the community reaches.

Results in Fig. 7 and Fig. 8 are reached under the hypothesis that the cells only make contact with their adjacent neighbours. Figure 9 shows how the community could increase the satisfaction by enabling the cells to contact more neighbours.

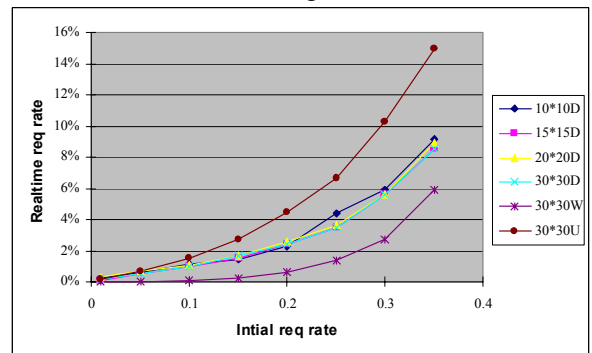

**Fig. 9 Increased contacts improve satisfaction**

In Fig. 9, all the simulations are carried out with 1% mutation rate. The curve with the highest req_rate value is the simulation of a 30×30 grid. The cells in this grid are only allowed to make access to their four adjacent neighbours. The four curves in Fig. 9 are grids with different size, they are all allowed to receive or provide help to four further neighbours additionally. The curve at the bottom is also a simulation of a 30×30 grid, with the ability to access further neighbours. But when counting the req_rate, this simulation only counts the cells that did not receive service; the cells that are being served are not accounted, while the other curves will record all the cells that need help. The results of these simulations show that enabling the cells to contact more neighbours would increase the satisfaction degree greatly. When the possibility to be a provider cell equals the possibility to be a client cell, the cells requiring for help in the community get reduced from 6% to 4%. If we do not exclude from this count those cells that are already receiving help, the number of waiting for help further drops to around 1%.

## 5. Conclusion

This paper investigated an Ambient Assisted Living System for elderly and disabled people. We developed a design tool to investigate the properties of an AAL System. The purpose is to let the AAL system automatically detect a cause of alarm, provide or accept help. Our AAL system is able to provide timely and cost-effectively service to elderly or disabled people, thus supporting their independent living in their houses.

In order to relief the work of the professional medical service, the system resorts for help on the healthy dwellers in the community. These dwellers are expected to occasionally provide non-medical service to their elderly or disabled neighbours. To incentivate their cooperation, the system could reward these "informal carers" both financially and morally.

The system proposed in this paper can be used to check how a mutual assistance community reacts to emergency situations, where the amount of individuals requesting help increases considerably while the pool of possible help providers consequently decreases.

Simulations in this paper showed that it is possible to provide the elderly and disabled with non-medical help by their neighbours promptly. The simulations also allowed us to reach the following conclusions:

1) If the states of the dwellers kept stable, the satisfaction of the client is only related to the percentage of providers. The percentage of client cells brings no impaction.

2) When the dwellers in the community are dynamically changing their states, the satisfactions of the clients are only related to the dweller's willingness to provide help and ask for help.

3) The size of the community did not bring direct effect on the satisfaction.

4) Increasing the number of neighbours that a dweller could make contact with will improve the satisfaction.

5) A community with higher stability, which means less dwellers mutating their states per time step, will bring higher satisfaction.

Further research with this model will develop the method of exchanging information in the form of message. In the model proposed in this paper, the way to get information of other individuals is by detecting the status and parameter values between neighbours. In our future research, we are going to design an ad hoc protocol to exchange information. The system will increase the contacts of a dweller from 8 neighbours to a broader scale.

In future research, experiments will be taken with the support of technological infrastructures. A "Living Lab" will be constructed to take experiments in reality. We define this "Living Lab" as an IT infrastructure that brings the various parties in contact through, e.g., wireless links, etc.

Besides the above researches on technological aspects, we plan to do research related to the involved psychological aspects. The role of the coordinator will be investigated and enhanced. The next stage research will also investigate different policies to improve service. To encourage the neighbours to provide their help, further research will also carried out to investigate the policies of rewarding the dwellers who provide help and introducing "penalties" for the dwellers who did not provide a promised service.

Communities like the ones discussed in this paper could also be run through a sort of membership fee. In this case reward could simply become discounts to the membership costs, while demerits could slightly increase costs, through principles similar to those used by insurance companies. Also these aspects shall be investigated.